# Carbon Nanoscrolls at High Impacts: A Molecular Dynamics Investigation

José Moreira de Sousa[1], Leonardo Dantas Machado[1,2], Cristiano Francisco Woellner[1], Pedro Alves da Silva Autreto[1,3], Douglas S. Galvao[1]

[1]Instituto de Física "Gleb Wataghin", Universidade Estadual de Campinas, Campinas - SP, 13083-970, Brazil

[2]Departamento de Física Teórica e Experimental, Universidade Federal do Rio Grande do Norte, Natal-RN 59072-970, Brazil

[3]Universidade Federal do ABC, Santo André-SP, 09210-580, Brazil

**ABSTRACT**

The behavior of nanostructures under high strain-rate conditions has been object of interest in recent years. For instance, recent experimental investigations showed that at high velocity impacts carbon nanotubes can unzip resulting into graphene nanoribbons. Carbon nanoscrolls (CNS) are among the structures whose high impact behavior has not yet been investigated. CNS are graphene membranes rolled up into papyrus-like structures. Their unique open-ended topology leads to properties not found in close-ended structures, such as nanotubes. Here we report a fully atomistic reactive molecular dynamics study on the behavior of CNS colliding at high velocities against solid targets. Our results show that the velocity and scroll axis orientation are key parameters to determine the resulting formed nanostructures after impact. The relative orientation of the scroll open ends and the substrate is also very important. We observed that for appropriate velocities and orientations, the nanoscrolls can experience large structural deformations and large-scale fractures. We have also observed unscrolling (scrolls going back to planar or quasi-planar graphene membranes), unzip resulting into nanoribbons, and significant reconstructions from breaking and/or formation of new chemical bonds. Another interesting result was that if the CNS impact the substrate with their open ends, for certain velocities, fused scroll walls were observed.

**INTRODUCTION**

The investigation of carbon-based materials at atomic scale has been subject of intense theoretical and experimental researches in recent years [1,2]. These studies have been mostly focused on the understanding the mechanical and electronic properties of these structures in quasi-static conditions. Only recently, extreme dynamic conditions were investigated through high velocity impact experiments which considered the nanostructures as target [3], as well as projectile.

Recent experimental/theoretical works showed that under certain conditions carbon nanotubes (CNT) can be unzipped when shot at high velocities against solid targets [4]. These investigations showed that the key parameters which determine the resulting structures from these collisions are the velocity values and the relative orientation of the CNT axis with relation to the target. These results stress the importance of the investigation of other nanostructures beyond quasi-static conditions.

A natural question is how other nanostructures behave under similar conditions. Among these nanostructures, graphene sheets rolled up into a papyrus-like form, the so-called carbon nanoscrolls (CNS) [5-9] (Figure 1), are of special interest. For a recent scroll review see [9]. In contrast to CNT, CNS have both end open, which provides them with a large radial flexibility (their diameters are easily tunable through physical and/or chemical processes) and large accessible surface area. Such properties can be exploited in applications such as nanoactuators [10] or hydrogen storage [11]. Scrolls of other materials, such as, boron nitride also exist [12].

In this work we have investigated the dynamical and structural properties of CNSs shot at high velocities against solid targets. We have considered cases that mimic the similar conditions of the CNT experiments [4].

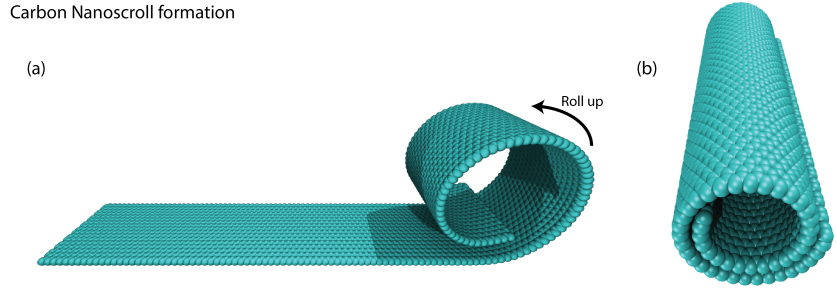

**Figure 1**: Scheme of the scroll formation through rolling up graphene membranes. (a) initial stage, and; (b) final stage.

**THEORY**

In order to investigate the structural, dynamical and mechanical properties of CNS under high impact (see Figure 2), we have carried out fully atomistic Molecular Dynamics (MD) simulations using the reactive force field ReaxFF [13], as implemented in the open source code LAMMPS [14,15]. The ReaxFF is a modern reactive force field which reproduces dissociation and/or breaking of chemical bonds of nanostructured systems. It is a potential that was developed to be a bridge between quantum and classical methods, where the parameters are obtained directly from first principles calculations and/or experiments [12]. Its relative low computational cost (in relation to *ab* initio methods, for instance) allows us to handle very large structures and long time simularions.

All MD simulations were carried out using a NVE ensemble. In order to avoid spurious simulation effects and to capture the physics characteristic of the fast collision phenomena, we used very small time steps (0.025 fs). We simulated high strain-rate conditions by shooting the nanoscrolls at ultrasonic velocities against a fixed van der Waals wall (rigid substrate).

The CNS used in our simulations have lengths in the order of 100 Å and are composed of ~ 4500 carbon atoms. In our simulations we considered impact velocity values ranging from 2 km/s up to 6 km/s, with increment values of 0.5 km/s. Besides the different velocity values, we have also considered different relative (in relation to the target) CNS orientations. Due to space limitations, we will discuss here only few cases (Figure 2):
- CNS collides with the substrate perpendicularly to its axis (vertical shooting in Figure 2), and;
- CNS collides with the substrate parallel to its axis (lateral shooting in Figure 2) for $\alpha=0$ and $90^o$.

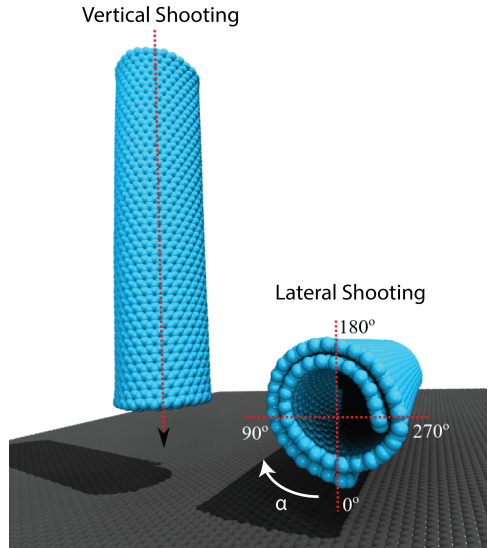

**Figure 2**: Schematic representation of the shooting cases (vertical and lateral shootings, respectively) of the nanoscrolls shot against a rigid substrate. The indicated α angle labeling refers to the scroll wall angle in relation to the substrate for lateral shooting.

**DISCUSSION**

In Figure 3 we present some representative cases obtained from MD simulations after CNS laterally impacting the substrate. Basically, we observed four main types of structures:

- **unscrolled structures (Figure 3a):** these structures occur when the impact does not produce structural collapses and/or fractures, but the acquired kinetic energy is large enough to overcome the van der Waals interactions (which tend to maintain the scrolled form) and the unscrolling (back to planar or quasi-planar graphene membranes) occurs;
- **closed ended structures (Figure 3b):** the closed ended case is a novel porous carbon nanostructure formed by partially welded (covalent bonds) CNS walls. It is an intermediate structure between CNT and CNS. The welded wall makes the radial expansion much more difficult, thus destroying the giant electrochemical effect reported to CNS [10];
- **bilayer nanoribbon** structures (Figure 3c): similarly to what was experimentally observed for CNT shooting [4], the CNS impact can also lead to bilayer carbon nanoribbon formation. This kind of structure is formed when the CNS impact the substrate and experiences elastic forces strong enough to flatten it. This flatness induces carbon-carbon breaking bonds, mainly around the folded regions. The fracture mechanisms and dynamics are quite similar to the CNT cases [4];
- **torn structures (Figure 3d):** these structures are formed under extreme conditions when the impact is so strong that most of the scroll structural integrity is destroyed. These kind of structures were also observed for CNT shootings [4].

Our results also showed that the formed structures are very dependent on the velocity and shooting orientations. A summary of the results for the obtained structures as a function of velocity and angle values is presented in Figure 4. Black rectangles represent cases in which the nanoscroll topology is preserved.

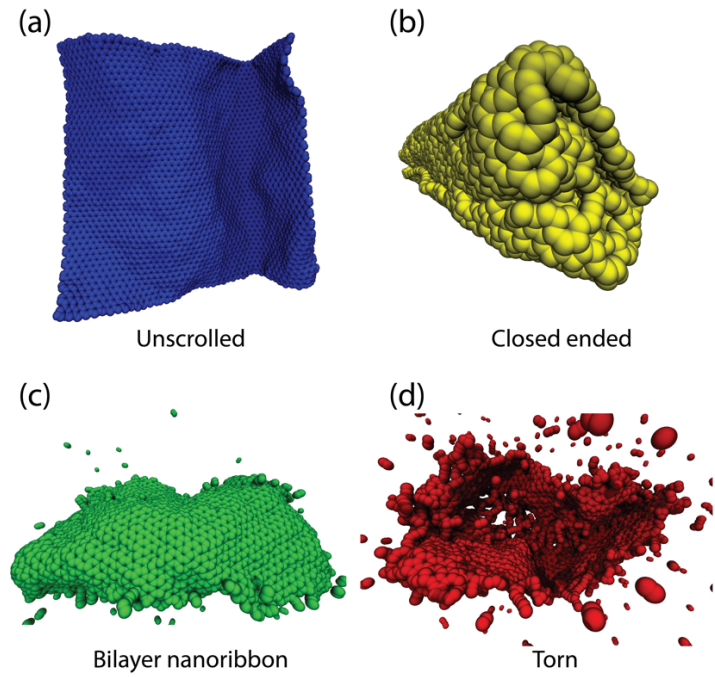

**Figure 3**: Representative formed structures after lateral impact collisions.

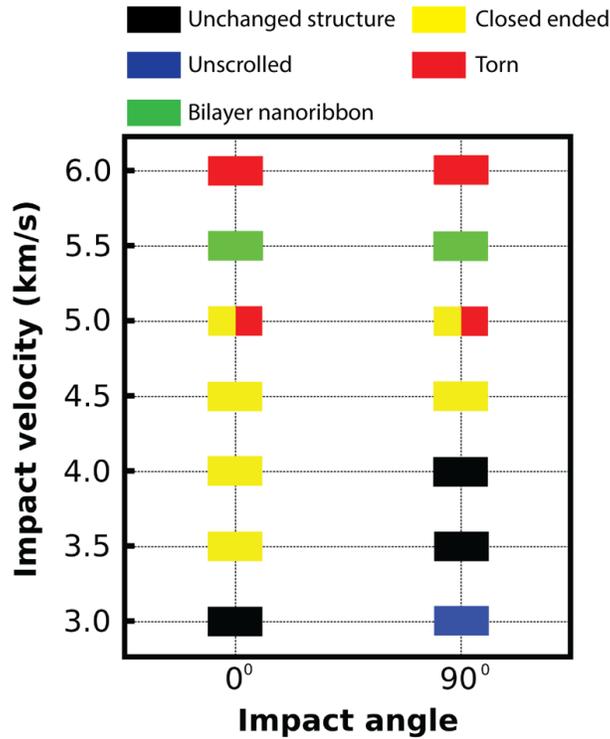

**Figure 4**: Summary of the kind of obtained structures as a function of impact velocity and angle values. Labeling corresponds to the structures shown in Figure 3. Dual color rectangles represent the cases where different structures coexist.

As can see from Figure 4 the relative orientation of the outlayer scroll (open edges) during impact is very important and directly affect the resulting structures after the impact. For the lowest velocity value investigated here (3 km/s) while the impact at 0° does not significantly affect the scroll conformation, at 90° induces unscrolling processes. This can be explained by the different degrees of freedom of the open edge (see Figure 2), constrained by the scroll itself and structurally free for 0° and 90°, respectively. Unscrolled structures were not observed for higher (> 3 km/s) velocity values, for any lateral impact angle.

The closed ended structures (partially welded walls) are mainly formed in the intermediate (from 3 up to 5 km/s) velocity values and at 0° where the open edges constrain favor the creation of covalent bonds between the scroll layers. For the 90° case, where the open edges have more free space to move, these structures are formed at a lower rate (see Figure 4).

For the bilayer structures, a perfect nanoribbon formation was only observed for 5.5 km/s. This velocity provides sufficient kinetic energy to flatten the CNS, breaking C-C bonds mainly on folded areas, while preserving the hexagonal structure of the graphene-like planes. For higher velocities (> 6 km/s), flatness also occurs but followed by significant structural damages (bond fractures), thus resulting in the torn structures.

For the vertical shootings, the structures remain scrolled in all cases. For velocities higher than 4 km/s evaporated and/or kinetically ejected carbon atoms started to be observed. In Figure 5, we present representative MD snapshots of the vertical shooting (2 km/s). Initially (indicated by I in the Figure 5) the tube undergoes a longitudinal deformation followed by a partial recovering (II), with small fractured regions at the edges. In Figure 5 it is also presented the kinetic and potential energies variation as a function of the simulation time. In contrast with the lateral shooting where part of the kinetic energy is dissipated through fracture and structural reconstructions, for the vertical shooting the kinetic energy is almost totally converted in potential energy through elastic deformations. It is also worth mentioning that the almost perfect elastic behavior is consequence of the scroll topology, this behavior is not observed for CNT where the impossibility of radial expansion lead to fractured structures.

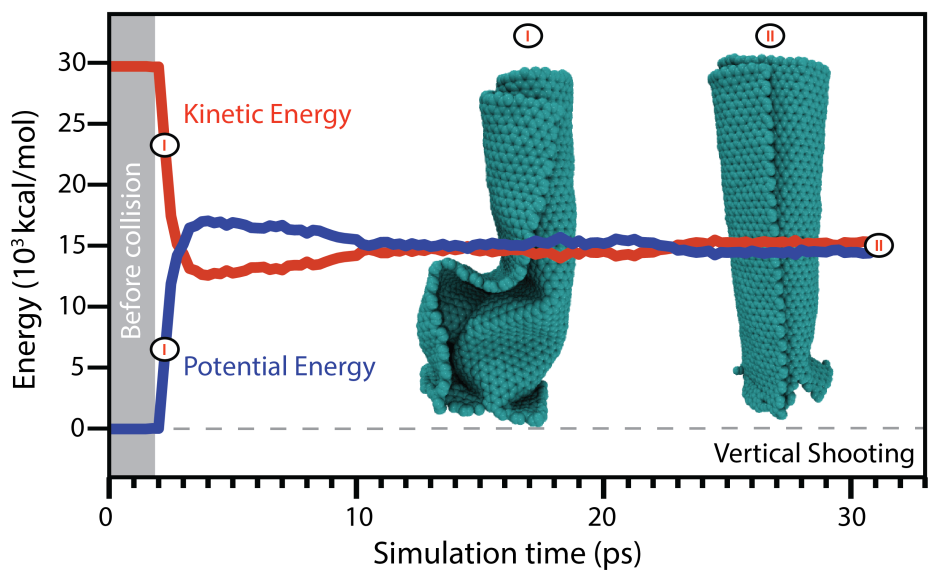

**Figure 5**: Kinetic and Potential energies for vertical shooting simulation as a function of simulation time.

## CONCLUSIONS

We have investigated through fully atomistic reactive molecular dynamics simulations the structural and dynamical aspects of carbon nanoscrolls (CNS) shot at high velocity values against solid targets. For the lateral shooting, depending on scroll velocity (from 2 up to 6 km/s) and relative orientation (in relation to the substrate) different (unscrolled, closed ended, bilayer nanoribbons and torn (highly fractured)_structures were observed. For the vertical shooting the scrolled topology survived for all the cases investigated here.

## ACKNOWLEDGMENTS

The authors acknowledge the São Paulo Research Foundation (FAPESP) Grant No. 2014/24547-1 for financial support. Computational and financial support from the Center for Computational Engineering and Sciences at Unicamp through the FAPESP/CEPID Grant No. 2013/08293-7 is also acknowledged.